\documentclass{optica-article}

\journal{opticajournal} 

\articletype{Research Article}

\usepackage{lineno}
\usepackage{subfig}
\usepackage{graphicx}
\usepackage{subcaption} 
\usepackage{booktabs} 

\begin{document}

\title{Suppressing data anomalies of gravitational reference sensors with time delay interferometry combinations}

\author{Pengzhan Wu,\authormark{1} Minghui Du,\authormark{1,*} and Peng Xu\authormark{1,2,3,4,\dag}}

\address{\authormark{1}Center for Gravitational Wave Experiment, National Microgravity Laboratory, Institute of Mechanics, Chinese Academy of Sciences, Beijing 100190, China\\
\authormark{2}Hangzhou Institute for Advanced Study, University of Chinese Academy of Sciences, Hangzhou 310024, China\\
\authormark{3}Taiji Laboratory for Gravitational Wave Universe (Beijing/Hangzhou), University of Chinese Academy of Sciences, Beijing 100049, China\\
\authormark{4}Lanzhou Center of Theoretical Physics, Lanzhou University, Lanzhou 730000, China}

\email{\authormark{*} Co-corresponding author: duminghui@imech.ac.cn \\
\authormark{\dag} Corresponding author: xupeng@imech.ac.cn} 


\begin{abstract*} 
For the LISA and Taiji missions, both transient and continuous data anomalies would pose significant challenges to the detection, estimation, and subsequent scientific interpretation of gravitational wave signals. 
As is indicated by the experiences of LISA PathFinder and Taiji-1, these anomalies may originate from the disturbances of the gravitational reference sensors due to routine maintenances and unexpected environmental or instrumental issues.
To effectively mitigate such anomalies and thereby enhance the robustness and reliability of the scientific outputs, we suggest to employ the ``position noise suppressing'' time delay interferometry channels.  
Through analytical derivations and numerical simulations, we demonstrate that these time delay interferometry channels can suppress data anomalies by more than 2 orders of magnitude within the sensitive band of 0.1 mHz - 0.05 Hz, while still remaining sensitive to most of the target signals.
Compared with existing researches that focus on reconstructing and subtracting data anomalies, our method does not rely on the  prior knowledge about the models of anomalies. 
Furthermore, the potential application scenarios of these channels have also been explored.
\end{abstract*}

\section{Introduction\label{sec:Introduction}}
The first detection of gravitational wave (GW) in 2015 by Adv-LIGO
\cite{the_ligo_scientific_collaboration_advanced_2015,abbott_observation_2016,abbott_observing_2016,abbott_properties_2016}, together with the subsequent observations by the LIGO-VIRGO collaboration in the following
years, had gradually open the new era of the gravitational wave astronomy.
To enclose the exciting sources in the millihertz band,
the pioneering mission concept of the Laser Interferometer
Space Antenna (LISA) was proposed in 1993 to the ESA's ``Horizon 2000 plus'' program~\cite{danzmann_lisa_1996,team_lisa_1997,danzmann_lisa_2000}.
Today, LISA, as the most fully-fledged mission concept under development, has been approved by ESA’s Science Programme Committee to build the instruments and spacecrafts, in preparation to be launched in 2035~\cite{danzmann_lisa_2017,lisa_red_book}.

Following the mission concept of ALIA~\cite{bender_additional_2004},
China had started her own pursuit of GW detection in space since 2008.
Then, under the collaboration between the Max Planck Institute for
Gravitational physics and the Chinese Academy of Sciences (CAS), the
first Chinese mission concept was proposed in 2011~\cite{gong_scientific_2011}
and afterward a more conservative design was made in
2015~\cite{gong_descope_2015,xue-fei_laser_2015}. Based on the preliminary studies, and also
encouraged by the breakthroughs made by both the LIGO and the LISA
PathFinder (LPF) mission~\cite{armano_sub-femto-_2016,armano_beyond_2018,armano_lisa_pathfinder_2019}, the ``Taiji Program in Space''
was released by the CAS in 2016 and the journey to China's space-borne
GW observatory had officially set forth~\cite{hu_taiji_2017,luo_taiji_2020,luo_brief_2020,ruan_taiji_2020}.
Taiji belongs to LISA-like missions, and according to its road map~\cite{luo_taiji_2020,luo_brief_2020},
it is expected that the science operations of LISA and Taiji may overlap in the 2030s. 
Studies of the space antenna network that formed by these two missions has now aroused more interest~\cite{omiya_searching_2020,ruan_lisataiji_2020,ruan_lisa-taiji_2021,wang_alternative_2021,wang_observing_2021}. 

For both the LISA and Taiji missions, the target GW sources, especially massive black hole binaries (MBHBs), galactic binaries (GBs), and extreme mass-ratio inspirals (EMRIs), produce GW signals that last for months or even years within the sensitive band. 
To precisely measure the waveforms and infer the physical properties of such sources, continues measurements without disruptions in the data streams are normally required. 
This imposes strong challenges on the long-term stability and robustness of both the high precision payloads and ultra-stable satellite platforms.
According to the designs of the LISA and Taiji missions, the foreseeable disruptions in science measurements may come from the scheduled maintenances and unexpected instrument anomalies. 
On the one hand, the scheduled maintenances include regular re-pointing of the telecommunication antennas, re-locking of lasers due to the switching of frequency plans~\cite{barke_simon_phdthesis,lisa_frequency_plan}, and orbital maneuvers~\cite{lisa_trajectory_design}, \emph{etc}. 
Scheduled maintenances could cause large and even continuous disturbances of satellite platforms and affect significantly the performances of the key payloads, such as the lasers, interferometers, and especially the gravitational reference sensors (GRSs), which are integrated to the movable optical sub-assemblies (MOSAs) to shelter the test-masses (TMs) and provide free-falling reference at the level of ${\rm 3 \times 10^{-15} \  m/s^2/\sqrt{Hz}}$ in the mHz band.
On the other hand, based on the experiences of the LPF~\cite{armano_sub-femto-_2016,armano_beyond_2018,baghi_detection_2022,lisa_pathfinder_collaboration_transient_2022}, 
GRACE~\cite{frommknecht_integrated_2007} and Taiji-1~\cite{the_taiji_scientific_collaboration_chinas_2021} missions, the GRSs or accelerometer systems could also be affected by unexpected anomalies manifested as, for example,  glitches in LPF's acceleration and interferometry readouts.

It is conservative to expect that the aforementioned GRS disturbances would take place during the science operations of both LISA and Taiji missions~\cite{baghi_preparation_2018,baghi_detection_2022}, leaving non-negligible anomalies in the interferometric data, and hence affect the detection and parameter estimation of GW signals. 
For example, glitches with low signal-to-noise ratios (SNRs) might be misinterpreted as burst-like GW signals~\cite{robson_detecting_2019}, while high-SNR glitches that overlap with MBHB mergers can severely damage the estimation of source parameters~\cite{Spadaro:2023muy}. 
If glitches occur repeatedly during the science operation, the overall noise level of the detector will also increase, as is the case with LPF~\cite{armano_lisa_pathfinder_2019}. 
Besides, the conventional matched filtering method for GW signal analysis is based on the assumption of Gaussian and stationary noise, however, the non-stationary and non-Gaussianity characteristics due to short-term glitches and long-term disturbances may require the employment of more complex likelihood functions~\cite{Sasli:2023mxr,baghi_gravitational-wave_2019}. 

Currently the physical origins of these data anomalies associated with GRS disturbances still remain to be fully explored. 
For LPF, the most possible explanation of glitches is due to the outgassing events  which occur in the spacecraft (S/C) about once a day~\cite{lisa_pathfinder_collaboration_transient_2022}. 
Considering the inheritance of technology and instruments, for the future LISA and Taiji missions, it is reasonable to infer that  GRS disturbances correlated with the S/Cs could take place in both GRSs on one of the three S/Cs. 
This is also the case when maneuver is applied to one of the S/Cs. 
In even more extreme cases, it would be possible that one of the S/Cs could not retain its ultra-stable and clean state for a rather long time due to some abnormal conditions, therefore the performances of GRSs onboard such S/C would be seriously affected and result into unwanted long-term data anomalies.

In face of these challenges, 
Q. baghi \emph{et al.} implemented the matching pursuit algorithm to detect and extract glitches from the  LPF data, using the shapelet functions as templates~\cite{baghi_detection_2022}. 
A. Spadaro \emph{et al.} conducted a joint estimation of modeled glitches and MBHB merger signals on simulated LISA data. 
Except for glitches, other realistic data anomalies such as non-stationary noises and gaps have also been taken seriously and drawn more attentions in recent studies~\cite{edwards_identifying_2020,dey_effect_2021}. 
When developing algorithms to cope with data anomalies, it should be kept in mind that the occurrence and morphology of disturbances in the future space-based GW detectors are still uncertain. 
Therefore, the applicability of aforementioned  methods in future scenario still  remains to be  verified. 
It would be ideal to find a model-independent method to effectively suppress these anomalies while still maintaining sensitivity to GW signals. 
{\color{black}Recent research efforts~\cite{AI1,AI2} have demonstrated the potential of artificial intelligence (AI) to detect and suppress a wide range of glitches and non-stationary noises. 
However, the reliability of neural networks usually needs to be  verified through more interpretable methods, which is also one of the motivations of our study.}

In this work, we reconsider the solution to data anomalies caused by the disturbances of GRSs. 
Different from present methods such as modelings and subtractions, we suggest a model-independent approach, 
namely the   ``position noise suppressing'' time delay interferometer (TDI) channels. 
This paper is arranged as follows. 
In Sec.~\ref{sec:Science-measurements}, we briefly review the measurement scheme shared by the LISA and Taiji missions, as well as the basic concepts of TDI. 
In Sec.~\ref{sec:Disturbances-from-GRS}, we summarize the forms of data anomalies associated with GRSs, according to the experiences of LPF, Taiji-1 and GRACE. 
In Sec.~\ref{sec: suppressing noise}, our ``position noise suppressing'' TDI channel is  theoretically derived and demonstrated through simulated data, and we have also predicted its potential application scenarios.
The concluding remarks of this paper are presented in Sec.~\ref{sec:Conclusions}. 
Finally, we provide the derivations of some important formulas in the appendix (Sec.~\ref{sec:appendix}).

\section{Measurement scheme and time delay interferometry}\label{sec:Science-measurements}

\begin{figure}
\centering\includegraphics[width=0.5\textwidth]{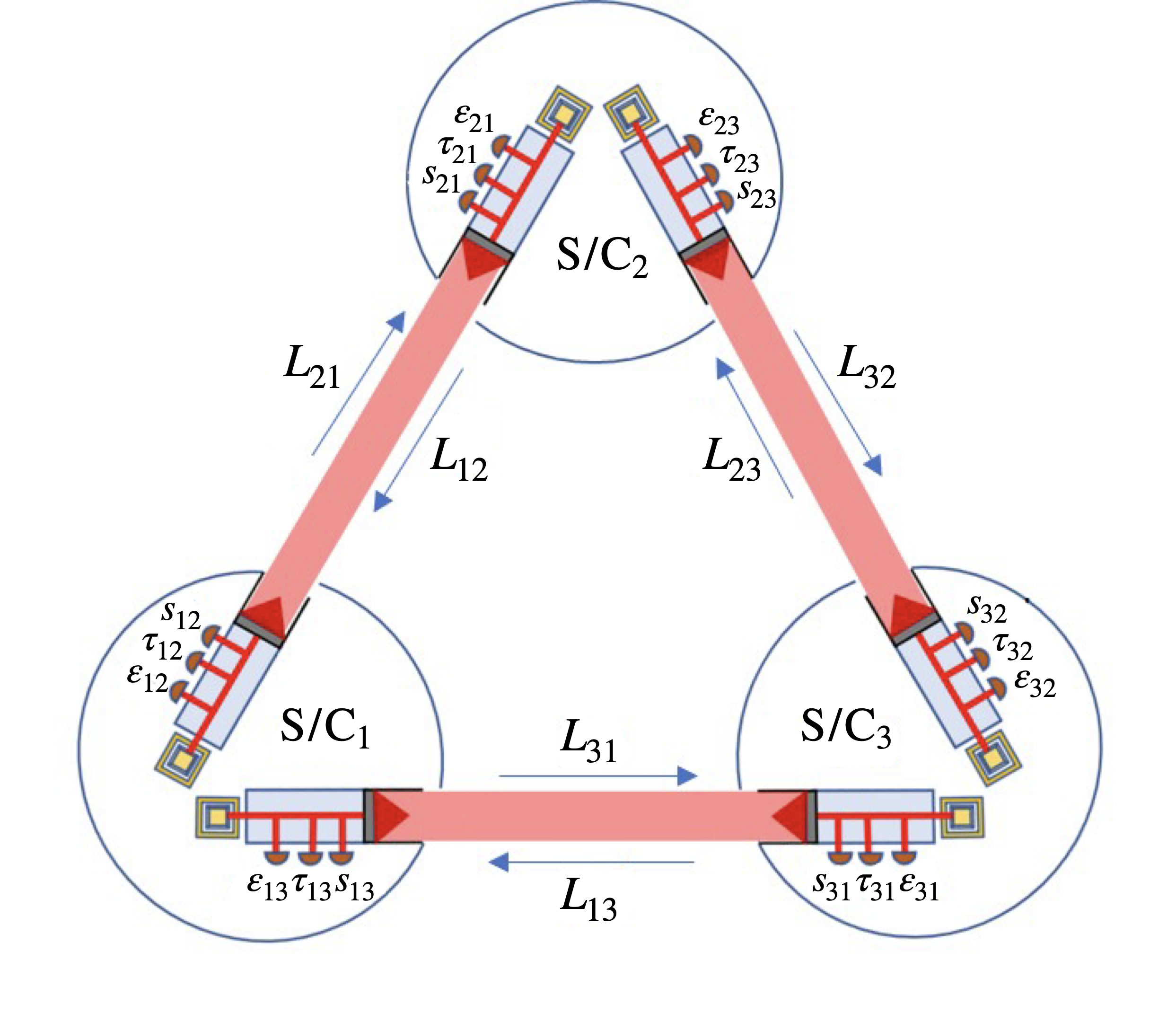}
\caption{The measurements scheme for the LISA and Taiji missions.}\label{fig:scheme}
\end{figure}

We introduce here the measurement scheme for the LISA and Taiji missions, and the notations used in the following sections. 
For both LISA and Taiji, the so-called ``split interferometer'' is adopted (see~\cite{otto_tdi_2012,otto_time-delay_2015,danzmann_lisa_2017} for detailed descriptions).  
Each one-way inter-spacecraft interferometry is divided into three parts: 
the inter-spacecraft science interferometer $s_{ij}(t)$ which links the optical benches (OBs) of S/C$_i$ and S/C$_j$; 
the local TM interferometer $\varepsilon_{ij}(t)$ that measures the distance between the local TM and the OB;  
together with the local reference (backlink) interferometer $\tau_{ij}(t)$ between the two local OBs of the same S/C.
Shown in Fig.~\ref{fig:scheme} are the labels of S/Cs, arms, OBs, and the aforementioned interferometers.  
Particularly, the optical path from S/C$_j$ to S/C$_i$ is denoted as $L_{ij}$. 
To facilitate the following descriptions, we  define 2 sets of indices: one is $\mathcal{I} \in \mathcal{I}_3$, representing $ij \in \{12, 23, 31, 21, 32, 13\}$ and $k$ (if present) is chosen so that $\{i, j, k\} = \{1, 2, 3\}$, and the other is $\mathcal{I} \in \mathcal{I}_3^{+}$, representing $ijk \in \{123, 231, 312\}$.

The readouts of the interferometers in the phase unit can be expressed as~\cite{bayle_simulation_2021,hartwig_instrumental_2021}
\begin{eqnarray}\label{eq:measurement_equation}
    s_{ij} &=& \boldsymbol{\rm D}_{ij} p_{ji} - p_{ij} + H_{ij} + \boldsymbol{\rm D}_{ij}\Delta_{ji} + \Delta_{ij} + N^s_{ij}, \nonumber \\ 
    \tau_{ij} &=& p_{ik} - p_{ij} + N^\tau_{ij}, \nonumber \\ 
    \varepsilon_{ij} &=& p_{ik} - p_{ij} + 2\Delta_{ij}  - 2\delta_{ij} + N^\varepsilon_{ij},
\end{eqnarray}
where $\mathcal{I}\in \mathcal{I}_3$. $p_{ij}$ denote the phase fluctuation caused by laser frequency instability, $\Delta_{ij}, \delta_{ij}$ the phase noises originating from the motions of OBs and TMs, and $N^s_{ij}, N^{\tau}_{ij}, N^{\varepsilon}_{ij}$ the optical metrology system (OMS) noises  of corresponding interferometers. The phase modulation  $H_{ij}$  caused by incident GWs is encoded in the inter-spacecraft science measurements $s_{ij}$. The time delay operator $\boldsymbol{\rm D}_{ij}$ is defined as $\boldsymbol{\rm D}_{ij} f(t) \equiv f(t-L_{ij}(t)/c)$, where $c$ is the speed of light.

The complete one-way inter-spacecraft interferometry $\eta_{ij}$ that connects the free-falling TMs can be then formed in post-processing following two steps, the first step being 
\begin{equation}\label{eq:intervar_xi}
    \xi_{ij} = s_{ij} + \frac{\tau_{ij} - \varepsilon_{ij}}{2} + \boldsymbol{\rm D}_{ij}\frac{\tau_{ji}-\varepsilon_{ji}}{2}, 
\end{equation}
where $\mathcal{I} \in \mathcal{I}_3$. And the second step is 
\begin{equation}\label{eq:intervar_eta}
    \eta_{ij} = \xi_{ij} + \boldsymbol{\rm D}_{ij}\frac{\tau_{ji}-\tau_{jk}}{2}, \quad \eta_{ik} = \xi_{ik} + \frac{\tau_{ij}-\tau_{ik}}{2},
\end{equation}
with $\mathcal{I} \in \mathcal{I}_3^+$. 
By inserting Eq.~(\ref{eq:measurement_equation}) into Eq.~(\ref{eq:intervar_xi}) and Eq.~(\ref{eq:intervar_eta}), the  OB noises and half of the laser noises are canceled out~\cite{otto_time-delay_2015,tinto_time-delay_2021}, thus 
\begin{equation}\label{eq:components_of_intervar_eta}
    \eta_{ij} = \boldsymbol{\rm D}_{ij}p_j - p_i + H_{ij} + \delta_{ij} + \boldsymbol{\rm D}_{ij}\delta_{ji} + N_{ij},
\end{equation}
where $\mathcal{I} \in \mathcal{I}_3$, and all the OMS noises are grounded in one term $N_{ij}$. 
The three left laser noises are $p_{12}, p_{23}, p_{31}$ and can be labeled with only 1 subscript, \emph{i.e.} $p_i \equiv p_{ij}$.

The one-way interferometry $\eta_{ij}$ is dominated by laser frequency noises, which are known as the primary noises in the science measurements of LISA-like missions, with magnitudes 6-8 orders beyond the expected sensitivity levels. 
To  mitigate  laser frequency noises, M. Tinto, F. B. Estabrook, and J. W.  Armstrong~\cite{tinto_cancellation_1999,armstrong_timedelay_1999}, W. T. Ni and collaborators~\cite{ni_progress_1997} in late 1990s had suggested to employ a post-processing technique  called TDI to construct virtual equal-arm interferometry from the aforementioned six one-way measurements. 
In the resulting TDI data streams (\emph{e.g.} Sagnac types, Michelson types),  the laser noises are  sufficiently suppressed, and  the secondary noises $\delta_{ij}, N_{ij}$ as well as GW signals $H_{ij}$ remain. 
A notable feature of  TDI  is that its capability to suppress laser  noise is independent of the functional form and statistical characteristics of noises, 
and only depends on the  mismatch between the synthesized virtual optical paths. 
The principle of TDI provides a heuristic idea for designing a model-independent method to  suppress data anomalies associated with GRS disturbances. 
Moreover, nowadays TDI acts as the basic framework of the initial noise reduction pipeline for LISA-like missions, therefore it is natural to consider the  issues of anomalies  within the TDI framework.   

\section{Data anomalies associated with  GRS disturbances\label{sec:Disturbances-from-GRS}}
\begin{figure}[ht!]
\centering
{\includegraphics[width=0.32\textwidth]{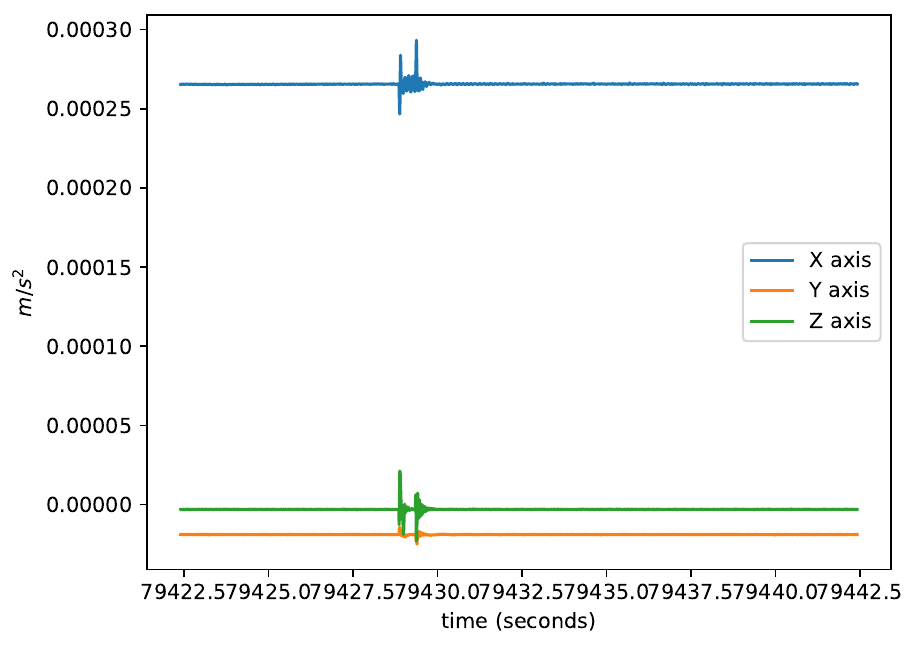}\label{fig:taiji-1-thruster}}
{\includegraphics[width=0.33\textwidth]{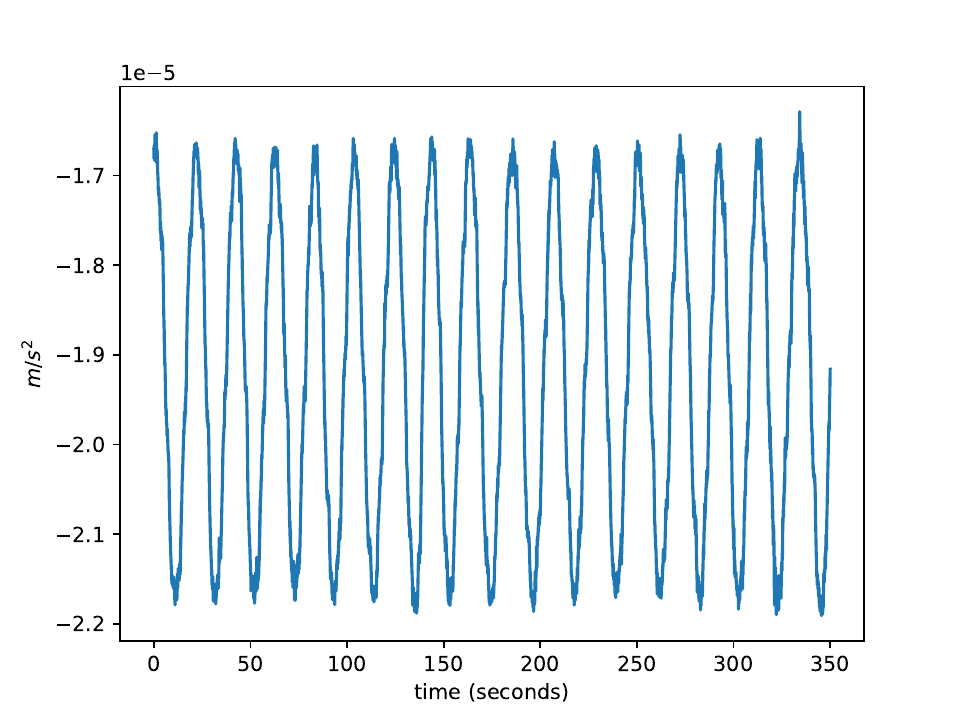}\label{fig:taiji-1-t}}
{\includegraphics[width=0.33\textwidth]{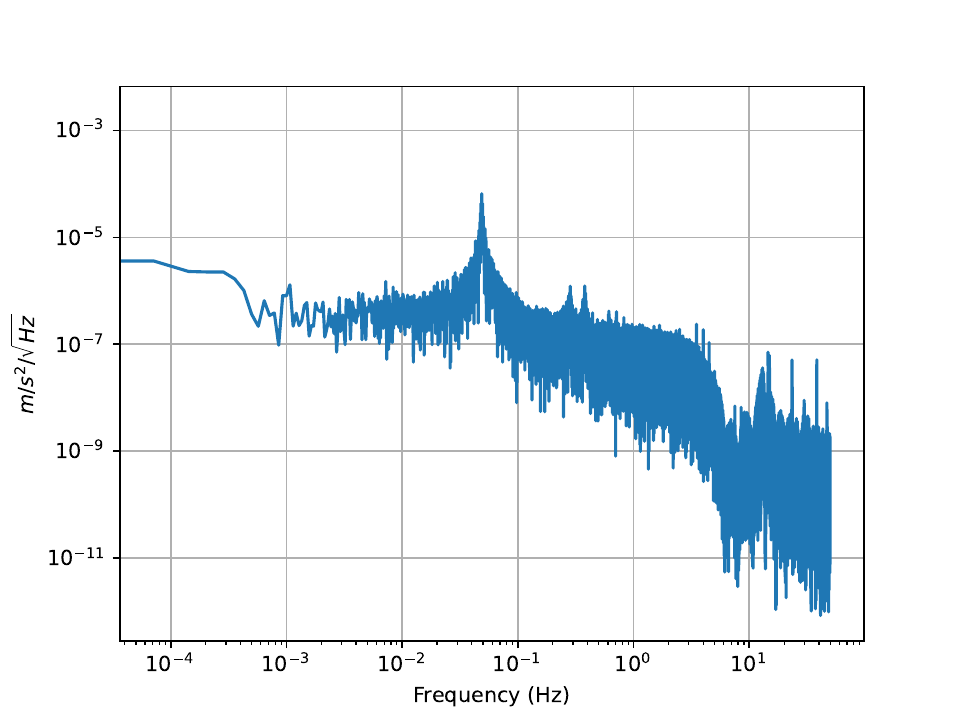}\label{fig:taiji-1-f}}
\caption{Typical responses of Taiji-1's GRS to small vibrations and continuous disturbances. Left panel: The responses of Taiji-1's GRS  to small vibrations caused by thruster events. Middle panel: Taiji-1's GRS readouts during the attitude adjustments with reaction wheels. Right panel: The ASD of Taiji-1's GRS readouts during the attitude adjustments with reaction wheels.\label{fig:Taiji-1_disturbances}}
\end{figure}

GRSs are one of the key payloads for LISA-like missions. The TMs suspended inside each GRS 
serve as the ultra-precision references of free-falling motions. 
Any perturbation to their free-falling states will produce noises that contaminate the science measurements.
For LISA and Taiji, the residual acceleration noises of the TMs along their sensitive axes are designed to be better than  $3\times 10^{-15} \ {\rm m/s^2/\sqrt{Hz}}$  in the mHz band~\cite{jennrich_lisa_2009,danzmann_lisa_2017,luo_taiji_2020,luo_brief_2020}. 
This means that the GRS systems are extremely sensitive to slight disturbances from payloads or platforms.
For example, the electrostatic accelerometers of GRACE, which can be seen as an alternative working mode of GRS, responded to events like the foil thermal effects and heating system switches, and produced transient phantom signals which had affected about 30\% of the measured data~\cite{frommknecht_integrated_2007,peterseim_twangs_2014}. 
To obtain a better performance, it is suggested to model and subtract such transients anomalies from the data in the pre-processing procedure~\cite{flury_precise_2008,peterseim_magnetic_2012,flechtner_identification_2014}. 
For LPF, unexpected  glitches were found to happen about once  per day during the science runs~\cite{baghi_preparation_2018,baghi_detection_2022,lisa_pathfinder_collaboration_transient_2022}. 
The requirement of the LISA free-falling performance could only be  achieved  after such glitches and other modeled noises were precisely removed from the data~\cite{armano_beyond_2018,baghi_detection_2022}. 
For the Taiji-1 mission~\cite{the_taiji_scientific_collaboration_chinas_2021,the_taiji_scientific_collaboration_taiji_2021}, the GRS was also found to respond to small vibrations like thruster events (see the left panel of Fig.~\ref{fig:Taiji-1_disturbances}). 

Besides, during the science run, we have also investigated how the GRS of Taiji-1 responded to large and continuous disturbances from the satellite platform, such as vibrations caused by reaction wheels during attitude adjustments~\cite{cai_satellite_2021,min_performance_2021,peng_system_2021,wang_development_2021}. 
In the middle and right panels of Fig.~\ref{fig:Taiji-1_disturbances}, one sees the $\sim 20 \ {\rm s}$ period signal in the time series and the  amplitude spectrum density (ASD) plots, which was due to the aliasing of the $\sim 6000 \ {\rm Hz}$  vibrations caused by reaction wheels. 
The overall noise level of the GRS had worsened during such maneuvers compared to the case of nominal science operations.  

Given the valuable experiences from these missions,  one could foresee that the performances of the GRSs for the LISA and Taiji missions may  be inevitably affected by 
unexpected environment disturbances, instrumental problems, and scheduled maintenances. 
In this paper, we take as examples the short-term glitches and  long-term continuous disturbances, which will manifest as  extra motions of TMs relative to the local free-falling inertial frames.
These extra motions will be picked up by the TM interferometers and appear in the TM-to-TM interferometry $\eta_{ij}$. Therefore, the total one-way interferometry reads
\begin{equation}\label{eq:intervar_eta_with_anomaly}
    \eta_{ij} = \eta_{ij}^{\rm E} + \eta_{ij}^{\rm N. T.}, \quad \eta^{\rm E}_{ij} = \delta^{\rm E}_{ij} + \boldsymbol{\rm D}_{ij}\delta^{\rm E}_{ij},
\end{equation}
where the superscript ``E'' stands for ``extra'' (\emph{i.e.} data anomalies), and ``N. T. '' denotes the ``normal'' terms shown in Eq.~(\ref{eq:components_of_intervar_eta}). 
Henceforth, the data anomalies $\eta^{\rm E}_{ij}$  will propagate through the TDI channels ~\cite{robson_detecting_2019,muratore_time_2022,hartwig_characterization_2022,wang_charactering_2022} and affect GW signal detections and parameter estimations, thus needs to be carefully accounted and resolved.  In the following sections, they will be collectively referred to as ``\textbf{GRS data anomalies}'', without causing any ambiguity.

In Sec.~\ref{subsec:numerical_simulation}, to provide a proof-of-principle demonstration, we simulate the aforementioned two classes of data anomalies. 
For glitches, according to LPF's observations, the most frequently occurred glitches are the fast rise and exponential decay type and the sine-Gaussian type~\cite{baghi_preparation_2018,baghi_detection_2022,lisa_pathfinder_collaboration_transient_2022}. 
We model  such transient anomalies  according to the ``LPF legacy model'',  as was adopted by the LISA Data Challenge (LDC)~\cite{lisa_data_challenge}. 
According to Taiji-1's experiments, the continuous disturbances in GRS caused by satellite maintenances can be modeled as the fast growths and gradually decays of the position noise $\delta_{ij}$ at the starts and the ends of the maneuvers respectively. 
The corresponding power spectral densities (PSDs) are hence time-varying. 
In principle, the morphology and statistical characteristics of  data anomalies could be much more complicated. While, as we aim to develop a model-independent anomaly suppressing method, these two models suffice to give a proof-of-principle demonstration. 

\section{Suppressing data anomalies with TDI channels \label{sec: suppressing noise}}

\subsection{Theoretical derivation}\label{subsec:theoretical_derivation}

As is mentioned before, when GRS data anomalies are recorded by  the one-way measurement $\eta_{ij}$, all the conventional TDI combinations which include the contributions from the disturbed S/C will be contaminated, and could not give rise to science data products retaining the expected $\mu {\rm rad/\sqrt{Hz}}$ noise level.
To resolve this problem, we introduce the ``position noise suppressing'' TDI channels $P_i^{(N)}$, with $i\in \{1, 2, 3 \}$ labeling the S/C where disturbances take place, and $N$ representing  the $N$th-generation  TDI. 

For simplicity, let's first focus on the 1st-generation TDI combinations. 
A good starting point would be  examining the GRS data anomalies in the 1st-generation Sagnac TDI channels. This is based on following considerations. Firstly, the Sagnac channels are widely  adopted and studied in the literature, which are proven to have sufficient capabilities to mitigate laser  noises  and competitive sensitivities  to detect GW signals. Secondly, they are known to generate the  solution space of all 1st-generation TDI channels~\cite{tinto_time-delay_2021,hartwig_characterization_2022}.  
The Sagnac-$\alpha, \beta, \gamma$ channels can be constructed from $\eta_{ij}$ as 
\begin{eqnarray}\label{eq:sagnac_channels}
    \alpha^{(1)} &=& \boldsymbol{\rm D}_{132}\eta_{21}+\boldsymbol{\rm D}_{13}\eta_{32}+\eta_{13} - \boldsymbol{\rm D}_{123}\eta_{31}-\boldsymbol{\rm D}_{12}\eta_{23}-\eta_{12}, \nonumber \\
    \beta^{(1)} &=& \boldsymbol{\rm D}_{213}\eta_{32}+\boldsymbol{\rm D}_{21}\eta_{13}+\eta_{21} - \boldsymbol{\rm D}_{231}\eta_{12}-\boldsymbol{\rm D}_{23}\eta_{31}-\eta_{23}, \nonumber \\
    \gamma^{(1)} &=& \boldsymbol{\rm D}_{321}\eta_{13}+\boldsymbol{\rm D}_{32}\eta_{21}+\eta_{32} - \boldsymbol{\rm D}_{312}\eta_{23}-\boldsymbol{\rm D}_{31}\eta_{12}-\eta_{31}, 
\end{eqnarray}
where the multiple-delay operator is defined as $\boldsymbol{\rm D}_{i_1i_2i_3...}\equiv \boldsymbol{\rm D}_{i_1i_2}\boldsymbol{\rm D}_{i_2i_3}...$. In general, the delay operators are not commutative (\emph{i. e.} $\boldsymbol{\rm D}_{i_1i_2}\boldsymbol{\rm D}_{i_3i_4}\neq \boldsymbol{\rm D}_{i_3i_4}\boldsymbol{\rm D}_{i_1i_2}$) due to the motion of S/Cs~\cite{PhysRevD.68.061303,Tinto:2003vj}. Assuming that all the GRSs are somehow disturbed, by inserting Eq.(~\ref{eq:intervar_eta_with_anomaly}) into Eq.~(\ref{eq:sagnac_channels}), we have 
\begin{eqnarray}\label{eq:anomalies_in_sagnac_channels}
    \alpha^{(1)} 
    &=& \left(\boldsymbol{\rm D}_{1321}-1\right)\delta^{\rm E}_{12} -\left(\boldsymbol{\rm D}_{1231}-1\right) \delta_{13}^{\rm E} + \left(\boldsymbol{\rm D}_{132}-\boldsymbol{\rm D}_{12}\right)\delta^{\rm E}_{\rm S/C_2} 
    -  \left(\boldsymbol{\rm D}_{123} - \boldsymbol{\rm D}_{13}\right)\delta^{\rm E}_{\rm S/C_3} \nonumber \\ 
    && + \ \alpha^{\rm (1) \ N.T.}, \nonumber \\
    \beta^{(1)} 
    &=& \left(\boldsymbol{\rm D}_{2132}-1\right)\delta^{\rm E}_{23} -\left(\boldsymbol{\rm D}_{2312}-1\right) \delta_{21}^{\rm E} + \left(\boldsymbol{\rm D}_{213}-\boldsymbol{\rm D}_{23}\right)\delta^{\rm E}_{\rm S/C_3} 
    -  \left(\boldsymbol{\rm D}_{231}-\boldsymbol{\rm D}_{21}\right)\delta^{\rm E}_{\rm S/C_1} \nonumber \\ 
    && + \ \beta^{\rm (1) \ N.T.}, \nonumber \\
    \gamma^{(1)}  
    &=& \left(\boldsymbol{\rm D}_{3213}-1\right)\delta^{\rm E}_{31} -\left(\boldsymbol{\rm D}_{3123}-1\right) \delta_{32}^{\rm E} + \left(\boldsymbol{\rm D}_{321}-\boldsymbol{\rm D}_{31}\right)\delta^{\rm E}_{\rm S/C_1} 
    -  \left(\boldsymbol{\rm D}_{312}-\boldsymbol{\rm D}_{23}\right)\delta^{\rm E}_{\rm S/C_2} \nonumber \\ 
    && + \ \gamma^{\rm (1) \ N.T.}.
\end{eqnarray}
Again ``N. T.'' denotes the ``normal'' terms including the secondary noises and GW signals (see Sec.~\ref{subsec:appendix2} for the detailed expressions), and we have defined the total anomaly onboard each  S/C:
\begin{equation}
    \delta^{\rm E}_{\rm S/C_1} = \delta_{12}^{\rm E}+\delta_{13}^{\rm E},\quad
    \delta^{\rm E}_{\rm S/C_2} = \delta_{23}^{\rm E}+\delta_{21}^{\rm E},\quad 
    \delta^{\rm E}_{\rm S/C_3} = \delta_{31}^{\rm E}+\delta_{32}^{\rm E}.
\end{equation}
As can be seen from Eq.~(\ref{eq:anomalies_in_sagnac_channels}), since the virtual optical path of $\alpha^{(1)}$ passes through the two TMs on S/C$_2$ (and S/C$_3$) simultaneously, data anomalies of the both GRSs onboard S/C$_2$ (and S/C$_3$) always appear together in the formula. Similar statements also applies to $\beta^{(1)}$ and $\gamma^{(1)}$,  making it possible to cancel $\delta^{\rm E}_{ij}$ and $\delta^{\rm E}_{ik}$ simultaneously using the  ``common-mode'' signals in other channels. 

According to  the experiences of LPF and Taiji-1, GRS data anomalies correlated to the S/Cs or platforms typically occur in both GRSs onboard a certain S/C, say S/C$_2$. Setting $\delta^{\rm E}_{13}, \delta^{\rm E}_{12}, \delta^{\rm E}_{32}$ and $\delta^{\rm E}_{31}$ to zero, we obtain 
\begin{eqnarray}
    \alpha^{(1)} 
    &=&  \left(\boldsymbol{\rm D}_{132}-\boldsymbol{\rm D}_{12}\right)\delta^{\rm E}_{\rm S/C_2}  + \alpha^{\rm (1) \ N.T.}, \\
    \gamma^{(1)}  
    &=& -\left(\boldsymbol{\rm D}_{312}-\boldsymbol{\rm D}_{23}\right)\delta^{\rm E}_{\rm S/C_2}  + \gamma^{\rm (1)  \ N.T.}.
\end{eqnarray}
A crucial observation is that the $\left(\boldsymbol{\rm D}^2 - \boldsymbol{\rm D}\right)\delta^{\rm E}_{\rm S/C_2}$ terms appear in  the expressions of $\alpha^{(1)}$ and $\gamma^{(1)}$ with negative  signs, thus  an anomaly-free data stream can be  constructed  as  
\begin{equation}
    P_2^{(1)} \equiv \frac{1}{\sqrt{2}} \left(\alpha^{(1)} + \gamma^{(1)} \right).
\end{equation}
A factor of $1/\sqrt{2}$ is introduced to balance the amplitude of secondary noises in the $P^{(1)}_2$ channel, making it  comparable to the original Sagnac channels. 
In the idealistic equal-arm case, the contributions of anomaly $\delta^{\rm E}_{\rm S/C_2}$ to the $\alpha^{(1)}$ and $\gamma^{(1)}$ channels can be perfectly cancelled. While due to the motions of S/Cs, there will be mismatch in their delay operators $\left(\boldsymbol{\rm D}_{132}-\boldsymbol{\rm D}_{12}\right)$ and  $\left(\boldsymbol{\rm D}_{312}-\boldsymbol{\rm D}_{23}\right)$, leaving uncancelled residuals in the $P^{(1)}_2$ channel. 
To provide a quantitative description, we define the suppression factor as
{\color{black}\begin{eqnarray}\label{eq:suppressing_factor_1st_generation}
    G_2^{(1)}(f) &\equiv& \frac{{\rm PSD \ of \ anomalies \ in \ }P^{(1)}_2}{{\rm PSD \ of \ anomalies \ in \ }\alpha^{(1)}} 
    = \frac{8 \cos^2 \left(u_{31}/2\right) \sin^2\left(\Delta u_2 / 2 \right)S^{\rm E}_{\rm S/C 2}}{4\sin^2 \left(\frac{u_{31}+\Delta u_2}{2}\right)S^{\rm E}_{\rm S/C 2}} \nonumber \\ 
    &\approx& \frac{2\cos^2(u/2)\sin^2(\Delta u_2/2) }{\sin^2(u/2) } 
    = 2 {\rm cot}^2\left(\frac{u}{2}\right) \sin^2\left(\frac{\Delta u_2}{2}\right), 
\end{eqnarray}}
with $u \equiv 2\pi f L / c$, $L$ being the nominal armlength, $u_{ij}\equiv 2\pi f L_{ij}/c$, $\Delta u_2 \equiv 2 \pi f \left(L_{23}-L_{21}\right) / c$, 
and $S^{\rm E}_{\rm S/C 2}$ is the PSD of $\delta^{\rm E}_{\rm S/C_2}$. At low frequencies ($\lesssim$ 0.05 Hz), $G^{(1)}_2(f)$ is approximately a constant $2(L_{23} - L_{21})^2/L^2$, while at high frequencies,  $G^{(1)}_2(f)$ oscillates around $\Delta u_2^2 / 2$. 
According to the optimized orbits of LISA and Taiji, the inequality of armlengths relative to the nominal armlength 
are $\lesssim 1\%$~\cite{PhysRevD.106.102005,Wang_2023}, thus normally a suppressing capability (characterized by the square root of $G$) of more than 2 orders   can be  achieved in the sensitive band 0.1 mHz - 0.05 Hz, which is sufficient even for loud glitches with SNRs up to  a few hundreds.

\begin{figure}[ht!]
    \centering
    \includegraphics[width=0.7\textwidth]{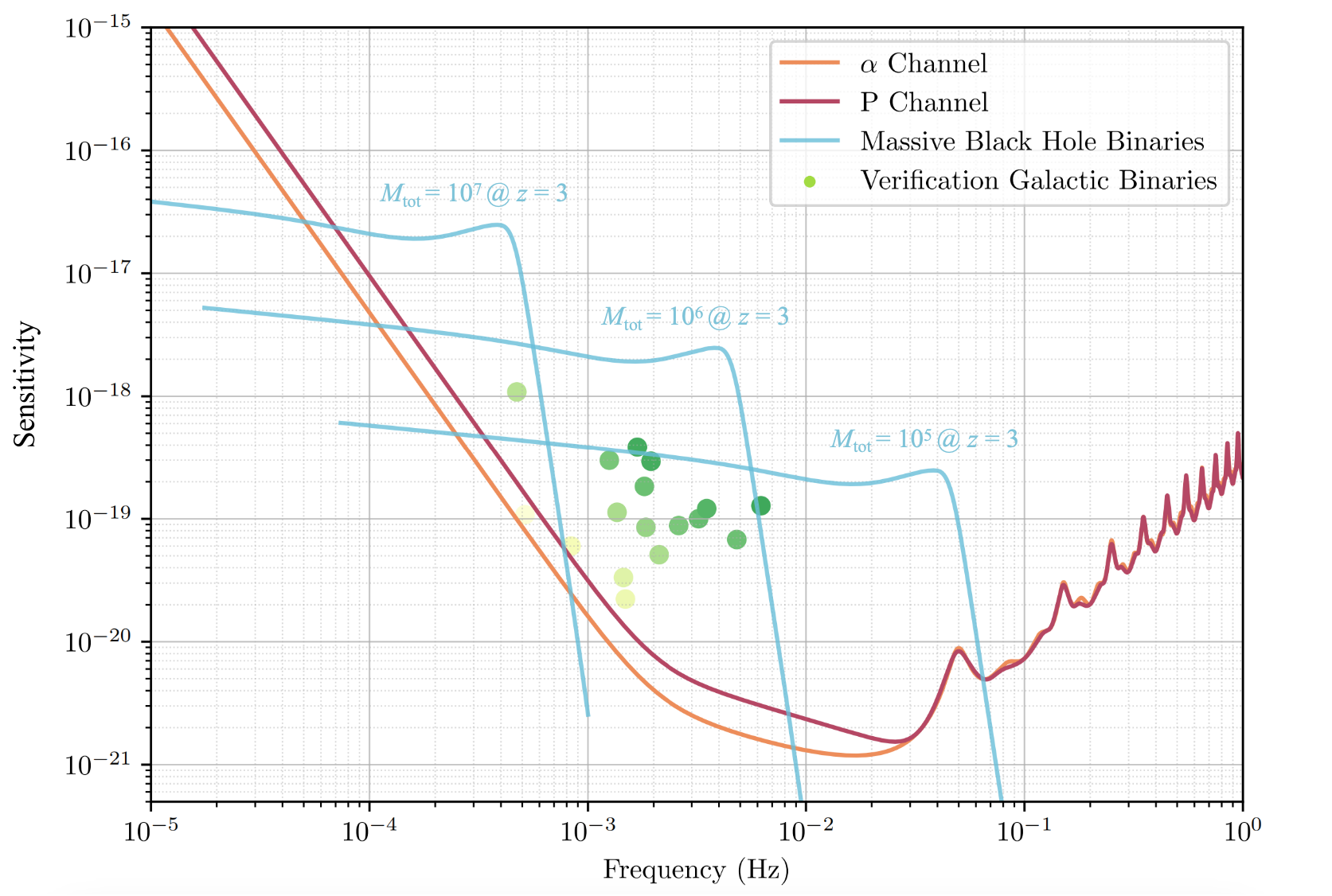}
    \caption{The sensitivity of  $P$ channel compared to the original Sagnac-$\alpha$ channel.}
    \label{fig:sensitivity}
\end{figure}

Following the same deviation, the position noise suppressing TDI channels for S/C$_1$ and S/C$_3$ are 
\begin{equation}
P_1^{(1)}\equiv \frac{1}{\sqrt{2}}\left(\beta^{(1)} +\gamma^{(1)}\right), \quad
P_3^{(1)}\equiv \frac{1}{\sqrt{2}}\left(\alpha^{(1)} +\beta^{(1)} \right). 
\end{equation}
Furthermore, this approach can be extended to TDI  schemes    beyond the 1st-generation.  
For the $N$th-generation case, we construct the position noise suppressing channels as
\begin{equation}
P_i^{(N)}=\frac{1}{\sqrt{2}}\sum_{j\neq i} \alpha_j^{(N)}, \label{eq:Pn}
\end{equation}
where $\alpha^{(N)}_j$ is the $N$th-generation Sagnac-type TDI combination, with $j \in \{1, 2, 3\}$ being the starting S/C of the virtual optical paths (e.g. $j=1, 2, 3$ for $\alpha^{(N)}, \beta^{(N)}, \gamma^{(N)}$, respectively). 
With similar algebra, 
we derive the GRS noise suppression factor $G^{(N)}_i$ to the leading order of armlength inequality:
\begin{equation}\label{eq:suppressing_factor_nth_generation}
    G^{(N)}_i(f) \equiv \frac{{\rm PSD \ of \ anomalies \ in \ }P^{(N)}_2}{{\rm PSD \ of \ anomalies \ in \ }\alpha^{(N)}}  
    \approx 2{\rm cot}^2\left(\frac{u}{2}\right) \sin^2 \left(\frac{\Delta u_i}{2}\right).
\end{equation}
where  $\Delta u_i \equiv 2\pi f (L_{ij} - L_{ik}) / c$. 
Comparing Eq.~(\ref{eq:suppressing_factor_1st_generation}) and Eq.~(\ref{eq:suppressing_factor_nth_generation}), the suppression factor does not explicitly depend on $N$, thus the order-of-magnitude analysis about $P^{(1)}_i$  naturally applies to $P^{(N)}_i$. 
An analytical proof of this statement can be found in Sec.~\ref{subsec:appendix1}. 
The suppression factor is visualized and compared with numerical  simulation in the right panel of Fig.~\ref{fig:suppress_factor}. 
The $P_i^{(N)}$  channels provide a model-independent approach  to  mitigate   the influences of GRS disturbances onboard S/C$_i$ by more than 2 orders in the sensitive band. 
Therefore, they may serve as  a new solution to  reduce  relevant risks for the LISA and Taiji missions. 
However, to be practically useful for data analysis, a TDI channel must not only reduce noises or anomalies, but also be sensitive to GW signals.  
Following the general definition  introduced in  Ref.~\cite{Robson:2018ifk}, we calculated the sensitivities of the $P$ channel and the original Sagnac channel for the Taiji mission. 
Shown in Fig.~\ref{fig:sensitivity} are the two sensitivity curves, as well as  several of the  typical  target sources. 
It appears that the $P$ channels are slightly less sensitive than the original Sagnac channels in the low frequencies, thus they may not be the optimal choices for  signal  extraction and parameter estimation during the regular science operations.  
However, when large glitches or long-term disturbances occur on S/C$_i$, the conventional Sagnac and Michelson TDI data streams will be contaminated.  
For example, as is reported in Ref.~\cite{Spadaro:2023muy}, using the conventional  TDI-AET channels, the overlapping of high-SNR glitches may cause the MBHB signals to be completely lost (\emph{i.e.} the estimations are more than 3$\sigma$s away from the truths). 
In these scenarios, 
the $P_i^{(N)}$ channels may become an indispensable choice, as they can ensure the detections and unbiased estimations of most signals without considerable lost of SNRs. 

Furthermore, the $P$ channels could have other potential applications. 
For instance, it may be used to distinguish between glitches and burst-like GW signals, through the comparison between  $P$ channels and  the conventional TDI channels. 
Additionally, by comparing the $P$ channels corresponding to different S/Cs, together with the auxiliary data of pressure, temperature, \emph{etc.}, the localization  of anomalies (\emph{i.e.} determining on which S/C the anomalies occur) might also be achieved. 
In the next section, several examples will be provided to demonstrate these applications.
 
\subsection{Numerical simulation}\label{subsec:numerical_simulation}

To demonstrate the performances and applications of the position noise suppression channels $P^{(N)}_i$, we have conducted two sets of  numerical simulations using  the simulation package  \texttt{TaijiSim}.  
Both sets are based on the  optimized numerical orbit of Taiji developed by Ref.~\cite{PhysRevD.106.102005}. 
For both LISA and Taiji, the 2nd-generation TDI can adequately suppress laser noises below the level of secondary noises~\cite{PhysRevD.68.061303,Tinto:2003vj,Cornish:2003tz}. 
Therefore, in this section, we focus on the case of $N = 2$. 
The duration of simulated data is set to $10^5$ s, with a sampling frequency of 1 Hz.

\begin{figure}[ht!]
\centering
{\includegraphics[width=0.32\textwidth]{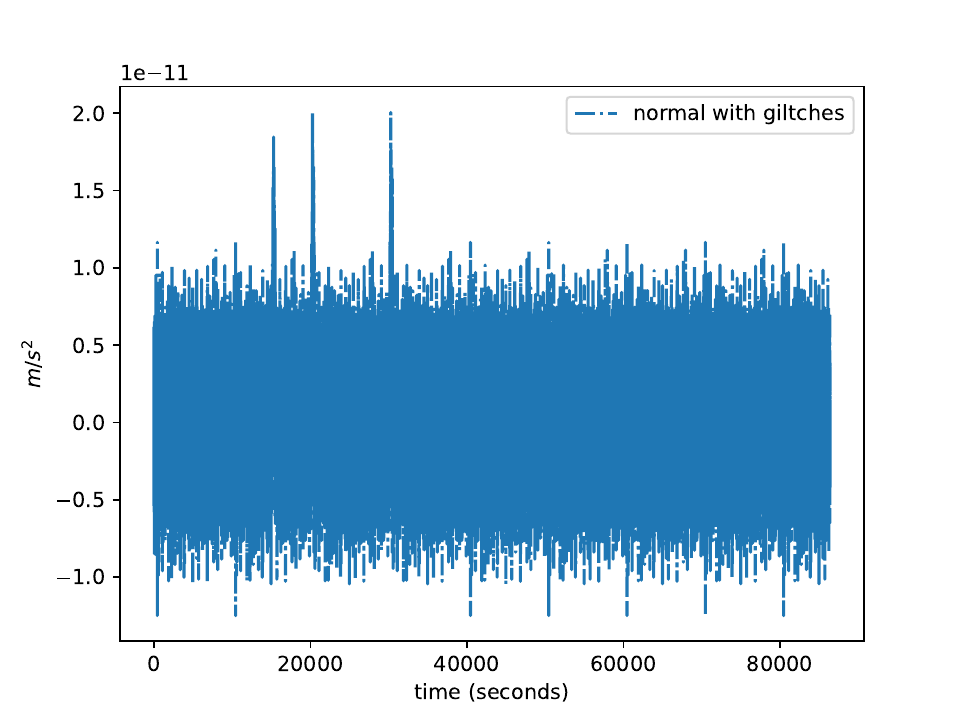}
\label{fig:glitch}}
{\includegraphics[width=0.33\textwidth]{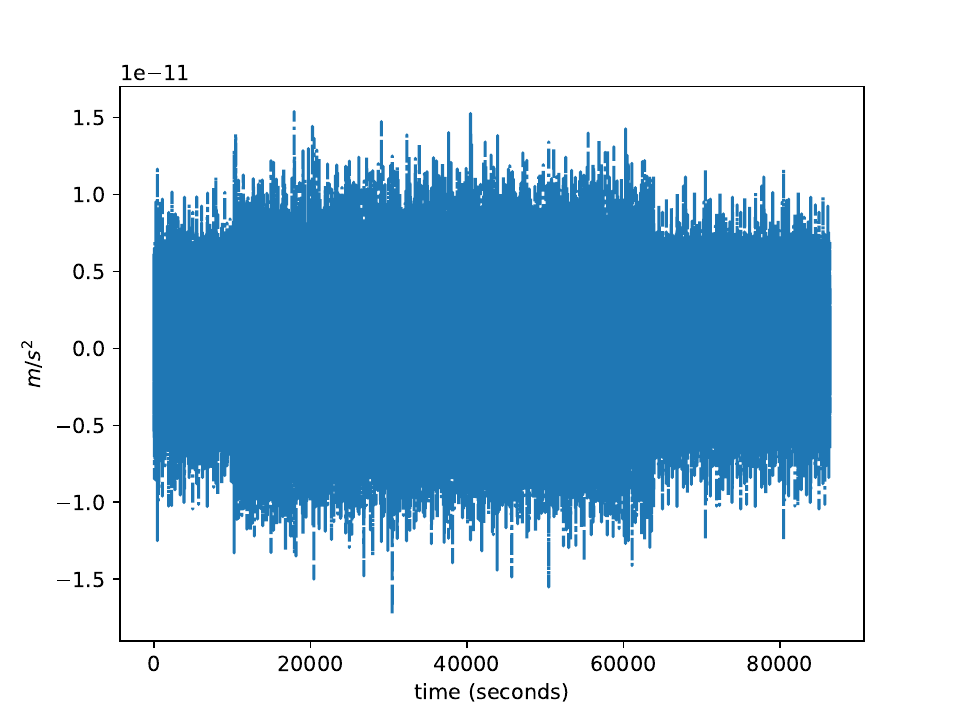}\label{fig:disturbance}}
{\includegraphics[width=0.33\textwidth]{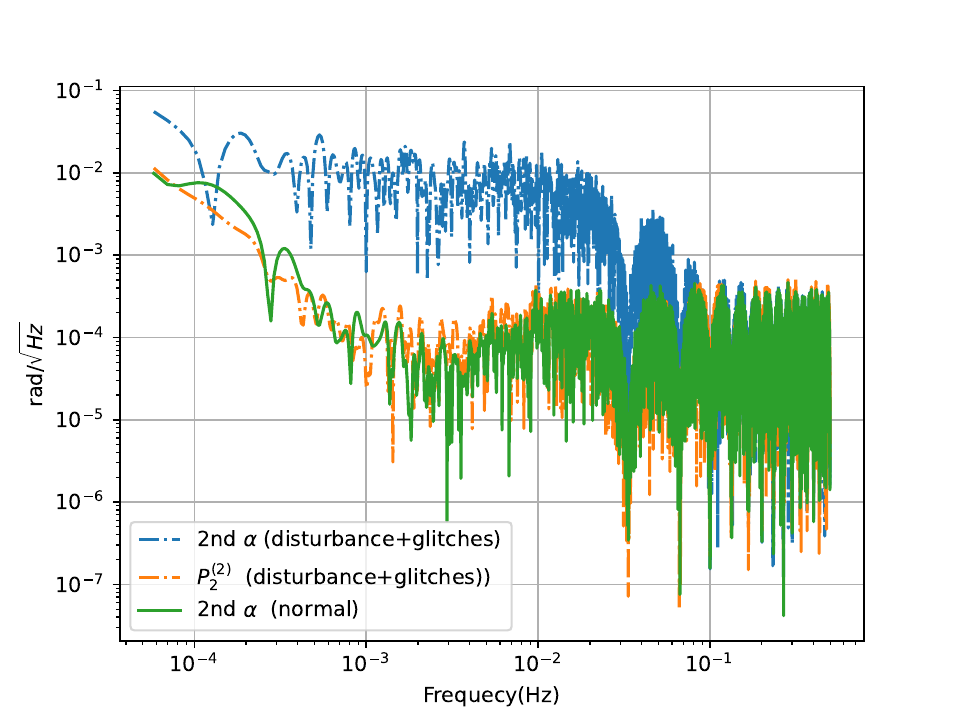}\label{fig:noiseasd}}
\caption{Simulations of TM acceleration   with ``normal'' noises and  different anomalies, and the resulting TDI channels. Left panel: Simulation of TM acceleration in GRS$_{21}$ with 3 glitches. Middle panel: Simulation of TM acceleration in GRS$_{21}$ with continuous disturbances. 
Right panel: ASDs of  Sagnac-$\alpha^{(2)}$ and $P^{(2)}_2$ TDI channels. The blue and orange curves correspond to the $\alpha^{(2)}$ and $P^{(2)}_2$ channels with data anomalies, and the green curve is  $\alpha^{(2)}$ channel with only ``normal'' noises.
}
\label{fig:simu-noise}
\end{figure}

The first set of simulations is designed to give a proof-of-principal demonstration on the function of $P$ channel to suppress various GRS data anomalies.
We generate a dataset with glitches, continuous disturbances, as well as the ``normal'' noises present in Eq.~(\ref{eq:measurement_equation}). 
All types of  anomalies are injected into the two GRSs onboard S/C$_2$.
The ``normal'' noises  are  simulated using the \texttt{TaijiSim-Noise} module. 
As is introduced in Sec.~\ref{sec:Disturbances-from-GRS},   the LPF legacy  model is adopted to simulate glitches, which has a double decaying exponential form in terms of acceleration:
\begin{equation}
    \delta^{\rm E}_{\rm acceleration}(t) = \frac{\Delta v}{\tau_1 - 
    \tau_2}\left(e^{-\frac{t-t_0}{\tau_1}} - e^{-\frac{t-t_0}{\tau_2}}\right)\Theta(t-t_0),
\end{equation}
where $t_0$ is the time of injection, $\Delta v$ is the velocity gain caused by the glitch, $\tau_{1,2}$ are the time scales of the exponentials, and $\Theta(t)$ stands for the Heaviside function. 
This model is integrated into the \texttt{TaijiSim-Glitch} module. 
As an example, the simulated acceleration data of GRS$_{21}$ with 3 randomly generated glitches  are visualized in the left panel of Fig.~\ref{fig:simu-noise}.
In the middle panel, we show the acceleration data of  GRS$_{21}$  with  continuous disturbance. 
The acceleration noise is increased  to  1.5 times the normal level during a time span of $\sim 5\times 10^4$ s. 
Both the   amplification of noise  and the edge effect will lead  to  the increase of overall noise level.   

Using the  \texttt{TaijiSim-TDI} module, how such GRS data anomalies affect the the ASDs of 2nd-generation TDI combinations are shown in the right panel of Fig.~\ref{fig:simu-noise}.  
As expected, when anomalies occur on S/C$_2$, they would significantly contaminate  the Sagnac-$\alpha^{(2)}$ channel (blue curve). 
Compared to the ``normal'' noise of $\alpha^{(2)}$ (green curve), an evident  increase due to the anomalies can be observed in the sensitive band. 
By using the  $P^{(2)}_2$ channel, the ``extra'' noises are sufficiently mitigated (orange curve), and the residual noises reduce to the normal level. 
As one could expect, with these channels the effects of the extra GRS disturbances are, in some sense, absent in the final data products.  

\begin{table}  
    \centering  
        \begin{tabular}{ccccccccccc}  
            \hline
            \hline
            $\mathcal{M}_c$ & $q$ &  $\chi_{z1}$ & $\chi_{z2}$ & $t_c$ & $\varphi_c$ & $d_L$ & $\iota$ & $\lambda$ & $\beta$ & $\psi$ \\
            \midrule  
            $5\times 10^5$ & 0.17 & 0.96 & -0.38  & 100 & 4.72 & $8.67\times 10^4$ & 1.80 & 2.76 & 1.46 & 0.30 \\ 
            \hline
            \hline
        \end{tabular}  \\
        \vspace{0.5cm}
        \begin{tabular}{cccc}  
            \hline
            \hline
            $\Delta v$ & $\tau_1$ & $\tau_2$ & $t_0$ \\  
            \midrule  
            $4.4\times 10^{-12}$ & 8.8 & 12.1 & 0 \\  
            \hline
            \hline
        \end{tabular} 
        \hspace{0.5cm}
        \begin{tabular}{cccc}  
            \hline
            \hline
            $\Delta v$ & $\tau_1$ & $\tau_2$ & $t_0$ \\  
            \midrule  
            $2.2\times 10^{-12}$ & 8.7 & 11.9 & 0 \\  
            \hline
            \hline
        \end{tabular} 
        \caption{Parameters of the injected MBHB merger signal and two glitches . The upper panel shows the parameters of MBHB. Order and units of the signal parameters: chirp mass $\mathcal{M}_c$ ($M_{\bigodot}$), mass ratio $q$, spin of the heavier black hole $\chi_{z1}$, spin of the lighter black hole $\chi_{z2}$, time of coalescence $t_c$ (s), phase at coalescence $\varphi_c$ (rad), inclination angle $\iota$ (rad), ecliptic longitude $\lambda$ (rad), ecliptic latitude $\beta$ (rad), polarization angle $\psi$ (rad). The lower two panels show the parameters of glitches injected into GRS$_{21}$ and GRS$_{23}$, respectively. Order and units of glitch parameters: $\Delta v$ (m/s), $\tau_{1,2}$ (s), $t_0$ (s).\label{tab:parameters_GW_glitches}}
\end{table}

The second set of simulations aims to quantitatively illustrate the capability  of  $P$ channel to suppress GRS data anomalies,  and showcase  how the $P$ channel responses differently to  anomalies and GW signals.
We simulated two datasets, one with a GW signal of MBHB merger, and  the other with two glitches  injected to GRS$_{21}$ and GRS$_{23}$ respectively. 
The GW signal is generated using the open-source software package \texttt{PyCBC}~\cite{PyCBC}, with the waveform template being SEOBNRv4~\cite{PhysRevD.95.044028}, and  their detector responses are calculated using the \texttt{TaijiSim-GW} module. 
The   glitch model is again the LPF legacy model. 
The parameters of GW signal and glitches are detailed in Table~\ref{tab:parameters_GW_glitches}. 
Notably, for GRS$_{21}$ and GRS$_{23}$, the strengths of glitches were set to twice and once of the ``short and loud'' glitch in the LDC spritz dataset, respectively. 

\begin{figure}[ht!]
    \centering
    {\includegraphics[width=0.48\textwidth]{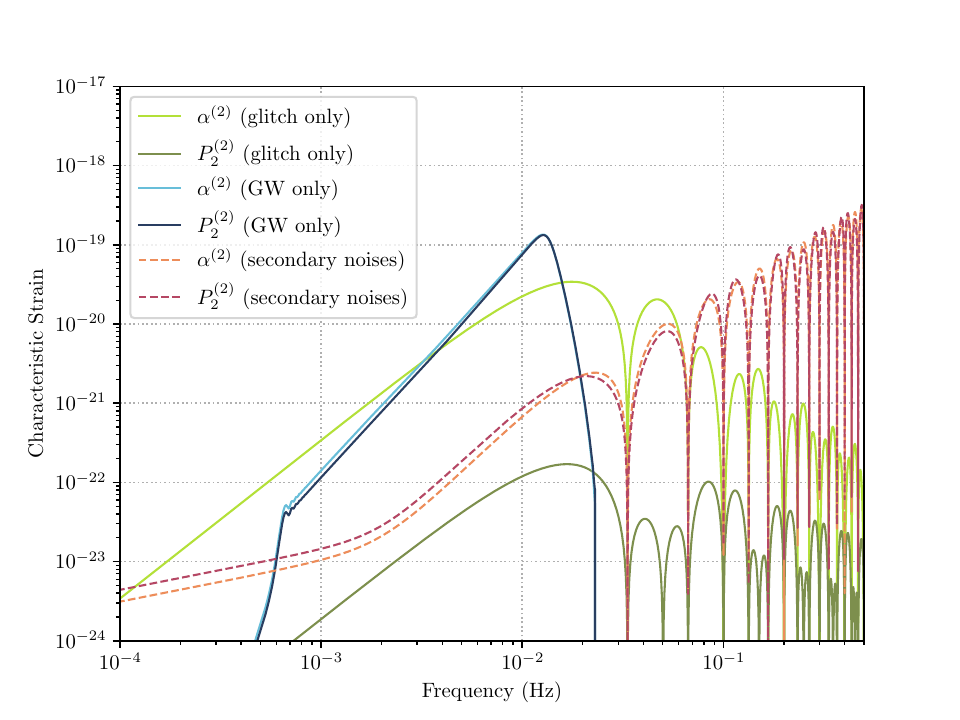}}
    {\includegraphics[width=0.48\textwidth]{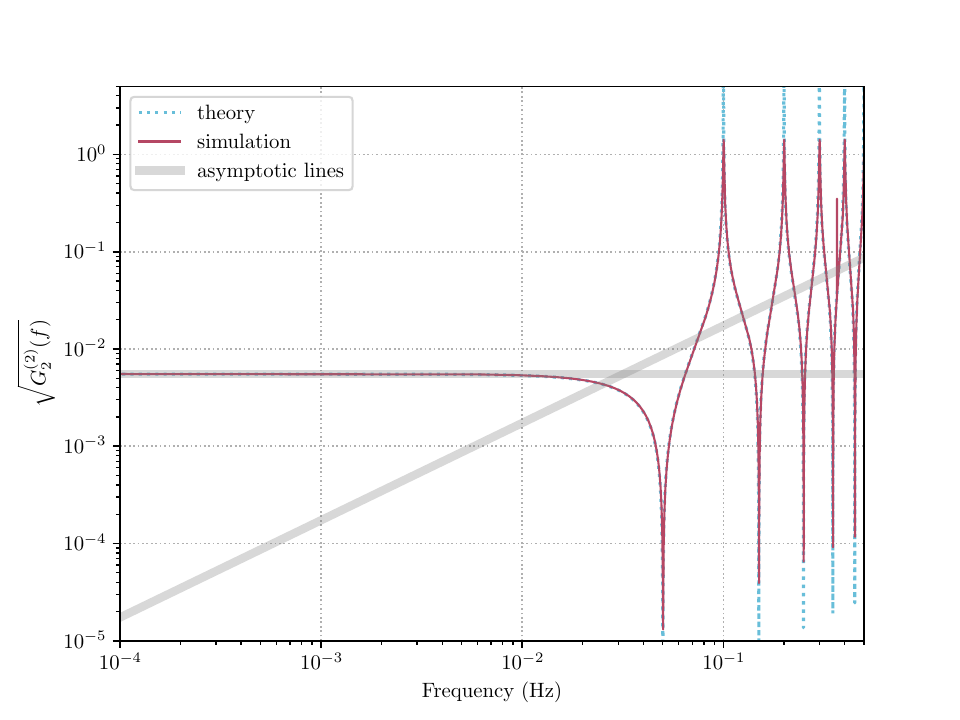}}
    \caption{Frequency-domain responses of the Sagnac-$\alpha^{(2)}$ and $P^{(2)}_2$ channels  to  glitches and GW signal. Left panel: Characteristic strains of the glitches (green curves) and GWs (blue curves) in the Sagnac-$\alpha^{(2)}$ and $P^{(2)}_2$ channels, compared to the theoretical secondary noises (red dashed curves). 
    Right Panel: Theoretical (blue dotted curve) and simulation (red curve) results of the suppression factor, with the asymptotic behaviour in the low and high frequencies shown with grey lines.}
    \label{fig:suppress_factor}
\end{figure}

We then construct the Sagnac-$\alpha^{(2)}$ and $P^{(2)}_2$ channels using \texttt{TaijiSim-TDI}. 
The left panel of Fig.~\ref{fig:suppress_factor} shows the characteristic strains of the two datasets in terms of different TDI channels, and the theoretical secondary noises  are also plotted for comparison (see Sec.~\ref{subsec:appendix2} for their expressions).
Visualizing the signals and glitches in terms of characteristic strains can provide an intuitive display of their SNRs. 
The total SNR of glitches in the $\alpha^{(2)}$ channel is 88.09, which is reduced to 0.36 after  suppressed by the $P^{(2)}_2$ channel, while there is only minor variation of SNR for the GW signal from $\alpha^{(2)}$ channel to $P^{(2)}_2$ channel (from 128.19 to 90.87). 
The right panel shows the square root of suppression factor $G^{(2)}_2$, namely the   amplitude ratio of anomaly between the $P^{(2)}_2$ channel and the $\alpha^{(2)}$ channel. The red curve represents the simulation result, the blue dotted curve indicates the theoretical result of Eq.~(\ref{eq:suppressing_factor_nth_generation}), and the grey lines shows the high-frequency and low-frequency asymptotic behaviors predicted in Sec.~\ref{subsec:theoretical_derivation}. 
The correctness of Eq.~(\ref{eq:suppressing_factor_nth_generation}) is verified through its consistency with the simulation.  
{\color{black}Eq.~(\ref{eq:suppressing_factor_nth_generation}) implies that the suppression factor exhibits infinities at $0.1n$ Hz and zeros at $0.1(0.5+n)$ Hz, while one should keep in mind that this is an approximate result, and for the general case without approximations (the 1st line of Eq.~(\ref{eq:suppressing_factor_1st_generation})), there are no infinities or zeros at these frequencies. }
Noting that even one of the injected glitch is twice as large  as the strongest one in the LDC spritz data,  a considerable suppression effect is still obtained, 
we hence conclude  that the $P$ channels have sufficient anomaly suppression capability in realistic scenarios. 
Moreover, these results clearly demonstrates the distinct responses of $P^{(2)}_2$ channel  to glitches and GW signals, thus one can also infer that  the $P$ channels  ensure reliable and accurate GW  parameter estimation in the presence of  overlapping  glitches. 

\begin{figure}[ht!]
\centering
{\includegraphics[width=0.9\textwidth]{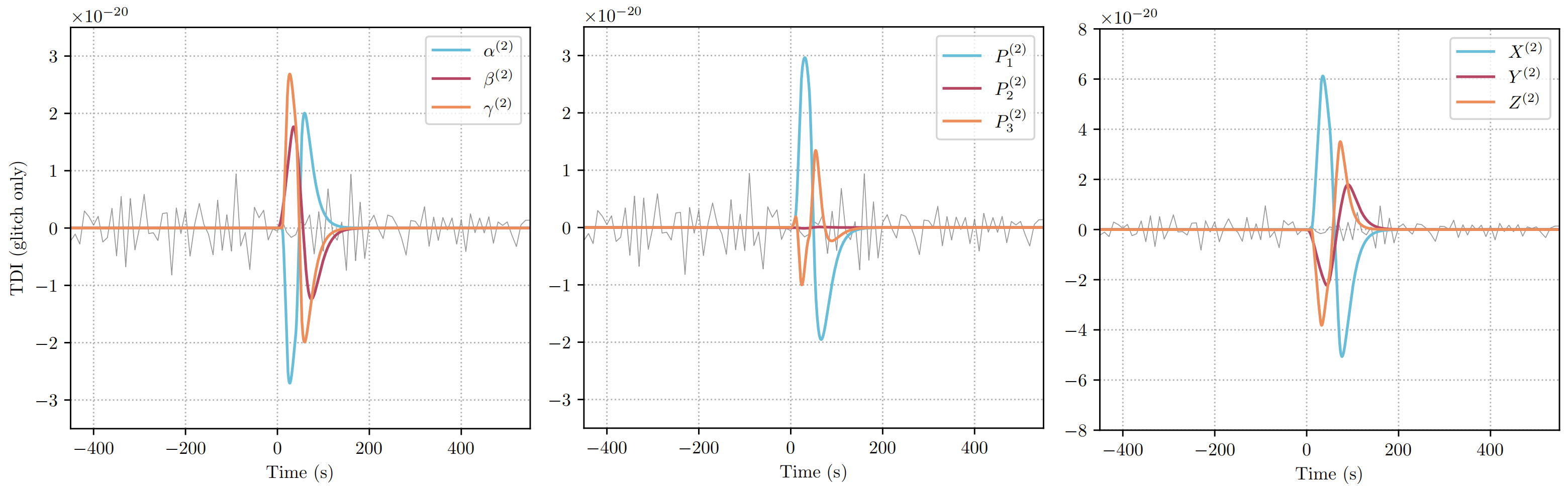}
\label{fig:TDI_glitch_example}} \\
{\includegraphics[width=0.9\textwidth]{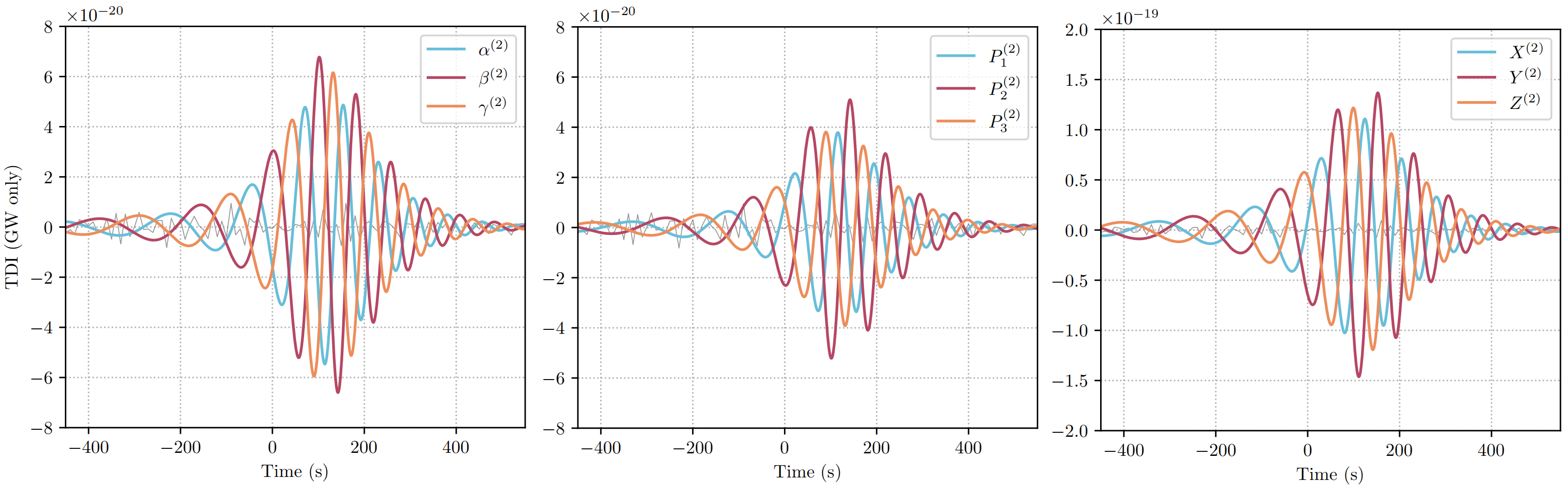}\label{fig:TDI_GW_example}} \\
\caption{Upper panels: Time-domain responses of the 2nd-generation Sagnac, $P$, and Michelson channels  to  glitches. Lower panels: Time-domain responses of the 2nd-generation Sagnac, $P$, and Michelson channels  to  GW signal. All the TDI data streams are converted to fractional frequency differences.}
\label{fig:TDI_example}
\end{figure}

Fig.~\ref{fig:TDI_example} displays the responses of different TDI channels to glitches and GW signals in the time domain, which might offer some insights into other practical applications of the $P$ channels. 
For comparison, we have also plotted the simulated ``normal'' noises with gray curves. 
All the TDI data streams are converted to the fractional frequency fluctuation unit, and down sampled to a sampling frequency of 0.1 Hz.  
According to Fig.~\ref{fig:suppress_factor}, this cut off is sufficient for the injected glitches and GWs.
{\color{black} Shown in the upper three panels are the ``glitch only'' dataset, where 
 $P^{(2)}_2$ exhibits a remarkable suppression effect on loud  glitches, while this suppression effect is absent for $P^{(2)}_1$ and $P^{(2)}_3$. 
This difference suggests that it might be possible to trace the locations of GRS data anomalies by comparing different $P$ channels. 
On the contrary, the amplitudes of glitches are similar in all the original Sagnac (or Michelson) channels, 
since they all include the contribution of $\delta_{{\rm S/C}_2}^{\rm E}$. 
The lower three panels show the ``GW only'' dataset. 
In the sense of morphology and amplitude, no significant change among  different channels can be observed. 
For the case in consideration, the Michelson channels demonstrate slightly higher sensitivities to the injected signal, while they are not robust against severe GRS data anomalies.
The comparison between the upper and lower panels indicates that, by jointly utilizing the $P$ channels together with other channels (Sagnac, Michelson, \emph{etc}), it might be possible to distinguish between unmodeled burst-like GW signals  and GRS glitches. }

\section{Concluding remarks\label{sec:Conclusions}}
For the LISA and  Taiji missions, data anomalies caused by the disturbances of GRSs due to  surrounding environments or the satellite platforms  will possibly take place during the science operations, and  pose challenges for the detection and estimation of GW signals. 

To resolve the issues of data anomalies, we suggest the employment of ``position noise suppressing'' TDI channels $P^{(N)}_i$, with $i \in \{1,2,3\}$ denoting the S/C where anomalies take place.  
By design such TDI channels can effectively suppress the data anomalies originating from the GRSs onboard S/C$_i$, without requiring any prior knowledge on the models of anomalies, whilst still remain sensitive to most of the target GW signals.   
With analytical derivations and numerical simulations, we have demonstrated and verified that the  anomalies  can be suppressed by  more than 2 orders  in the sensitive band 0.1 mHz - 0.05 Hz, given the optimized mission orbits with armlength inequality of $\lesssim 1 \%$. 
These channels may become an indispensable choice during the presence of severe  GRS disturbances. 
Furthermore, other potential applications of the $P$ channels have also been explored.
In summary, the TDI channels introduced in this paper significantly diminish the risks associated with GRS systems, thereby ensuring the robustness of scientific outcomes.

\section{Appendix}\label{sec:appendix}

\subsection{Derivation of suppressing factor for the $N$th-generation TDI}\label{subsec:appendix1}
In this appendix, we prove our statement analytically  that the GRS anomaly suppression capability of the $P$ channels is irrelevant to the generation of TDI. 

As is shown in Sec.~\ref{subsec:theoretical_derivation}, the $\delta^{\rm E}_{\rm S/C_2}$ terms in the $\alpha$ and $\gamma$ channels can be completely canceled out when all the armlengths of are equal and constant.  
However, several factors may prevent the perfect cancellation of anomalies. 
Firstly, according to the optimized orbital design of the LISA and Taiji missions, the inequality in armlengths are typically $\lesssim$ 1\% of the nominal armlength. 
Secondly, the inequality in the round-trip laser propagation time is a 0.5-order post-Newtonian correction, 
and is $\sim 10^{-4}$ of the nominal armlength.
Thirdly, within the time interval involved in the combination of TDI, the armlengths can vary with time. For the case of $N < 5$, these changes are typically less than $10^{-6}$ of the nominal armlength.
During the derivation of  suppression factor, we will only keep to the leading order terms, \emph{i.e.} the contributions from the first factor, and ignore the sub-leading ones.  
Thus it follows that $\boldsymbol{\rm D}_{ij} = \boldsymbol{\rm D}_{ji}$,  and all the delay operators are commutative.
Note that these approximations are only valid in the derivation of $G^{(N)}_i$. When implementing the TDI algorithm to suppress laser noises, all  the delays  should be calculated strictly according to the  numerical orbits.


The $N$th-generation  $\alpha$ channel can be expressed in a unified form: 
\begin{equation}\label{eq:Nth_generation_Sagnac}
    \alpha^{(N)} = \boldsymbol{\rm F}^{(N)}_{\alpha +} \alpha_{+} - \boldsymbol{\rm F}^{(N)}_{\alpha-} \alpha_{-},
\end{equation}
where $\alpha_+$ (or $\alpha_-$) stand for the terms with plus signs (or minus signs) in the first line of Eq.~\ref{eq:sagnac_channels}, 
and  the $\boldsymbol{\rm F}$ operators are defined as  
\begin{eqnarray}\label{eq:F_operators}
    && \boldsymbol{\rm F}^{(1)}_{\alpha +} = \boldsymbol{\rm F}^{(1)}_{\alpha-} = 1, \nonumber \\ 
    && \boldsymbol{\rm F}^{(2)}_{\alpha +} = 1 - \boldsymbol{\rm D}_{1231}, \quad \boldsymbol{\rm F}^{(2)}_{\alpha-} = 1 - \boldsymbol{\rm D}_{1321}, \nonumber \\
    && \boldsymbol{\rm F}^{(3)}_{\alpha +} = 1 - \boldsymbol{\rm D}_{1231} + \boldsymbol{\rm D}_{1321231}, \quad  
    \boldsymbol{\rm F}^{(3)}_{\alpha-} = 1 - \boldsymbol{\rm D}_{1321} + \boldsymbol{\rm D}_{1231321}, \nonumber \\ 
    && \boldsymbol{\rm F}^{(4)}_{\alpha +} = \left(1 - \boldsymbol{\rm D}_{1231}\right)\left(1 + \boldsymbol{\rm D}_{1321231}\right), \quad  \boldsymbol{\rm F}^{(4)}_{\alpha-} = \left(1 - \boldsymbol{\rm D}_{1321}\right)\left(1 + \boldsymbol{\rm D}_{1231321}\right), \nonumber \\ 
    && ... 
\end{eqnarray}
Based on above approximation,  it  holds that $\boldsymbol{\rm D}_{1231} = \boldsymbol{\rm D}_{1321}$, hence all $\boldsymbol{\rm F}^{(N)}_{\alpha+} = \boldsymbol{\rm F}^{(N)}_{\alpha-}$. 
On the other hand, for the $\gamma^{(N)}$ channels, according to the permutation rule of indices $(1\rightarrow 2, 2\rightarrow 3, 3 \rightarrow 1)$, $\boldsymbol{\rm D}_{1231}$  should be replaced by $\boldsymbol{\rm D}_{3123} = \boldsymbol{\rm D}_{31}\boldsymbol{\rm D}_{12}\boldsymbol{\rm D}_{23} = \boldsymbol{\rm D}_{12}\boldsymbol{\rm D}_{23}\boldsymbol{\rm D}_{31} = \boldsymbol{\rm D}_{1231}$, indicating that the $\boldsymbol{\rm F}$ operators in the $\gamma^{(N)}$ channels are essentially the same as   $\alpha^{(N)}$. 
Consequently, in the expressions of $\alpha^{(N)}$, $\gamma^{(N)}$,  and hence $P^{(N)}_2$, the $\boldsymbol{\rm F}$ operators can be factored out. 
Further, the noise transfer functions resultant from $\boldsymbol{\rm F}$s are also common for  $P^{(N)}_2$ and $\alpha^{(N)}$, and can be cancelled out in the ratio $G^{(N)}_2$. 
Similar deductions also apply to other $P^{(N)}_i$ channels. 
In summary, to the leading order of armlength inequality, the suppression factors are irrelevant to the generation of TDI. 


\subsection{The ``normal'' terms in TDI combinations}\label{subsec:appendix2}
For the sake of completeness, we also present here the expression of the ``normal'' terms in the  TDI combinations. 
For the original 1st-generation Sagnac channels,
\begin{eqnarray}\label{eq:normal_terms_in_alpha}
    \alpha^{\rm (1) \ N.T.} &=& \boldsymbol{\rm D}_{132}H_{21}+\boldsymbol{\rm D}_{13}H_{32}+H_{13} - \boldsymbol{\rm D}_{123}H_{31}-\boldsymbol{\rm D}_{12}H_{23}-H_{12} \nonumber \\
    && + \ \boldsymbol{\rm D}_{132}N_{21}+\boldsymbol{\rm D}_{13}N_{32}+N_{13} - \boldsymbol{\rm D}_{123}N_{31}-\boldsymbol{\rm D}_{12}N_{23}-N_{12} \nonumber \\
    && + \ \left(\boldsymbol{\rm D}_{1321}-1\right)\delta_{12} -\left(\boldsymbol{\rm D}_{1231}-1\right) \delta_{13} + \left(\boldsymbol{\rm D}_{132}-\boldsymbol{\rm D}_{12}\right)\left(\delta_{23} + \delta_{21}\right) \nonumber \\ 
    && - \ \left(\boldsymbol{\rm D}_{123} - \boldsymbol{\rm D}_{13}\right)\left(\delta_{31} + \delta_{32}\right).
\end{eqnarray}
Again, with the permutation rule of indices, one can obtain the expressions for $\beta^{\rm (1) \ N.T.}$ and $\gamma^{\rm (1) \ N.T.}$, and Sagnac channels beyond the 1st-generation can be deduced using Eq.~(\ref{eq:Nth_generation_Sagnac}) and Eq.~(\ref{eq:F_operators}). 

The characteristic strains of secondary noises for $\alpha^{(2)}$ and $P^{(2)}_2$ channels are plotted in the left panel of Fig.~\ref{fig:suppress_factor}, which are defined as~\cite{Robson:2018ifk}
\begin{equation}
    h_{n, {\rm TDI}}(f) \equiv \sqrt{f S_{\rm TDI}(f)},
\end{equation}
$S_{\rm TDI}(f)$ being the PSD of secondary noises in a certain TDI channel. 
To calculate $S_{\rm TDI}(f)$, we omit GW signals (\emph{e.g.} the 1st line of Eq.~(\ref{eq:normal_terms_in_alpha}), and take the autocorrelations  of the remaining terms. For $\alpha^{(2)}$, 
\begin{equation}
    S_{\alpha^{(2)}}(f) = 8 \sin^2
    \frac{u_{1231}}{2} \left[3 S_N(f) + 4 S_\delta(f) \left(\sin^2\frac{u_{1231}}{2} + \sin^2 \frac{u_{132} - u_{12}}{2} + \sin^2\frac{u_{123} - u_{13}}{2}\right)\right],
\end{equation}
and for $P^{(2)}_2$, 
\begin{equation}
    S_{P^{(2)}_2}(f) = \frac{1}{2}\left[S_{\alpha^{(2)}}(f) + S_{\gamma^{(2)}}(f) + 2 S_{\alpha^{(2)}\gamma^{(2) *}}(f)\right], 
\end{equation}
where 
\begin{equation}
    S_{\alpha^{(2)}\gamma^{(2) *}}(f) =  8 \sin^2 \frac{u_{1231}}{2} \left\{S_N(f)\left(2\cos u_{13} + \cos u_{123}\right) + 2 S_\delta(f) \left[\cos u_{13} - \cos (u_{23} - u_{12}) \right]\right\},
\end{equation}
with $u_{i_1i_2i_3...}\equiv 2\pi f (L_{i_1i_2} + L_{i_2i_3} + ...) / c$.  
$S_N(f)$ and   $S_\delta(f)$ are the PSDs of OMS noise and acceleration noise, respectively, and for them we adopt the models of Refs.~\cite{wang_algorithm_2021,luo_taiji_2020}.

\begin{backmatter}
\bmsection{Funding}
This work is supported by the National Key Research and Development Program of China No. 2020YFC2200601, No. 2021YFC2201901, No. 2021YFC2201903, and the
Strategic Priority Research Program of the Chinese Academy of Sciences
Grant No. XDA1502110202-02 and No. XDA1502110104-02. 






\bmsection{Disclosure}
The authors declare no conflicts of interest.

\bmsection{Data Availability Statement}
Data underlying the results presented in this paper are not publicly available at this time but may be obtained from the authors upon reasonable request.

\end{backmatter}

\bibliography{sample}






\end{document}